\documentclass[twoside]{article}
\usepackage{fleqn,espcrc2,epsfig}
\input epsf.sty

\newcommand{\nwc}{\newcommand}

\nwc{\be}  {\begin{equation}}
\nwc{\ee}  {\end{equation}}
\nwc{\bmu} {\bar{\mu}}
\nwc{\ba}  {\begin{eqnarray}}
\nwc{\ea}  {\end{eqnarray}}
\nwc{\bi}  {\begin{itemize}}
\nwc{\ei}  {\end{itemize}}
\nwc{\nn}  {\nonumber\\}
\nwc{\Tr}  {\mathop{\rm Tr}}
\nwc{\im}  {\mathop{\rm Im}}
\nwc{\re}  {\mathop{\rm Re}}
\nwc{\tb}  {\tan\!\beta}
\nwc{\Hc}  {\mathop{\rm H.c.}}
\nwc{\eq}  {Eq.~}
\nwc{\eqs}  {Eqs.~}
\nwc{\fig} {Fig.~}
\nwc{\la}[1]{\label{#1}}
\nwc{\rmi}[1]{{\mbox{\scriptsize #1}}}
\nwc{\nr}[1]{(\ref{#1})}
\nwc{\fr}[2]{{\frac{#1}{#2}}}
\nwc{\msbar}{\overline{\mbox{\rm MS}}}

\def\lsi{\raise0.3ex\hbox{$<$\kern-0.75em\raise-1.1ex\hbox{$\sim$}}}
\def\gsi{\raise0.3ex\hbox{$>$\kern-0.75em\raise-1.1ex\hbox{$\sim$}}}
\nwc{\lsim}{\mathop{\lsi}}
\nwc{\gsim}{\mathop{\gsi}}

\title{Spontaneous CP violation on the lattice%
       \thanks{Work partly supported by the TMR network 
       {\em Finite Temperature Phase Transitions in Particle
       Physics}, 
       contract no.\ FMRX-CT97-0122.
       Presented  
       by M. Laine.}}

\author{M. Laine\address{Theory Division, CERN, CH-1211 Geneva 23,
        Switzerland} 
        and 
        K. Rummukainen\address{NORDITA, Blegdamsvej 17,
        DK-2100 Copenhagen \O, Denmark}} 

\begin{document}

\begin{abstract}
At finite temperatures around the electroweak 
phase transition, the thermodynamics 
of the MSSM can be described by a three-dimensional two Higgs
doublet effective theory. This effective theory has a phase where 
CP is spontaneously violated. We study spontaneous CP violation with 
non-perturbative lattice simulations, and analyse whether one could 
end up in this phase for any physical MSSM parameter values. 
\end{abstract}
\maketitle

\vspace*{-10.3cm}

\begin{minipage}[t]{15.5cm}
\begin{flushright}
\mbox{CERN-TH/99-262} \\
\mbox{NORDITA-1999/51HE} \\
\mbox{hep-lat/9908045} \\
\mbox{August, 1999}
\end{flushright}
\end{minipage}

\vspace*{7.4cm}

\section{INTRODUCTION}

The physics problem we address is CP violation 
around the finite temperature MSSM electroweak phase transition. 
CP violation is one of the requirements for
baryogenesis, and it is thus of interest to ask 
whether it would be available to a sufficient extent in the 
MSSM, where the phase transition can be 
strong enough to account for the non-equilibrium 
constraint~\cite{bjls,mssmsim}. 

In particular, we are here interested in the phenomenon of 
spontaneous CP violation. Spontaneous CP violation can 
in principle take place in any two Higgs doublet model~\cite{lee},
but for the physical MSSM parameters it cannot be 
realized at $T=0$~\cite{mp}.  However, 
it has been suggested that it might be more easily realized 
at finite temperatures~\cite{emq}, or even only in the 
phase boundary between the symmetric and broken
phases~\cite{cpr,fkot}. 

As described 
in~\cite{cpown}, this questions can 
be studied with 3d effective 
fields theories and non-perturbative lattice simulations. 
More specifically, we report here on preliminary 
results concerning the following two questions: \\
1. Is there a homogeneous
thermodynamical phase in the MSSM
for some parameter values and temperatures, where
CP is spontaneously violated? \\
2. Could CP also be violated
at the phase boundary between the usual symmetric and
CP conserving broken phases? \\

\section{EFFECTIVE THEORY}

Let $O_a = H_a^\dagger H_a$, $a=1,2$, and 
$M = H_1^\dagger \tilde H_2$, $\tilde H_2 = i \sigma_2 H_2^*$. 
The theory we consider here is
\ba
& & \hspace*{-0.5cm} {\cal L}_\rmi{3d} =   
\fr14 F^a_{ij}F^a_{ij} 
+ \sum_{a=1}^{2} \Bigl[ (D_i H_a)^\dagger(D_i H_a) \nn
& & \hspace*{-0.5cm} 
+  m_a^2 O_a
+ \lambda_a O_a^2\Bigr] 
+ \lambda_3 O_1O_2  
+ \lambda_4 M^\dagger M  \la{action}\\
& & \hspace*{-0.5cm} +  \Bigl[m_{12}^2 M + \lambda_5 M^2 + 
 (\lambda_6 O_1+ 
 \lambda_7 O_2) M + \Hc\Bigr]. \nonumber
\ea
This theory is an effective theory for finite $T$ MSSM if
the right-handed stop mass is not smaller than  
the top mass. The opposite case leading to a strong transition,
necessitates a dynamical SU(3) stop triplet field, 
but we defer that discussion to a future publication ---  
the results do not differ qualitatively from the present 
ones. We have also neglected 
the dynamical effects of the hypercharge U(1) subgroup.

Using the symmetries of the Lagrangian, we can parametrize
the two Higgs doublets as 
\be
H_1 = \frac{v_1}{\sqrt{2}}
\left( 
\begin{array}{l}
1 \\
0
\end{array}
\right), \quad
\tilde H_2 = \frac{v_2}{\sqrt{2}}
\left( 
\begin{array}{l}
\cos\theta e^{i\phi} \\
\sin\theta
\end{array}
\right). \la{prms}
\ee
It turns out that in the case
relevant to us, $\lambda_4-2\lambda_5 < 0$, the angle 
$\theta$ settles dynamically to zero, so that we do not 
consider it any more. 

For real parameters, this theory is even  
under both of the discrete symmetries C, P. 
The C symmetry, $H_a \to H_a^*$, corresponds
to $\phi\to -\phi$. While parity is not
spontaneously broken in this theory, the C symmetry
can be, thus violating also CP~\cite{lee}.
CP violation is signalled by $|\cos\phi|<1$.

\section{TREE-LEVEL CRITERIA}

The region of the parameter space where there is 
CP violation can be found by minimizing the effective potential
of the theory in \eq\nr{action} with respect to $\phi$. We do this
first at the tree-level.

First of all, it is easy to understand that to get $|\cos\phi|<1$, 
one needs
\be
\lambda_4 - 2 \lambda_5 < 0,\quad\lambda_5 > 0.
\ee
The first condition is automatically satisfied in, 
say, the MSSM
where $\lambda_4 \approx -g^2/2$. The second condition 
can also be satisfied:  $\lambda_5>0$ (as well as
$\lambda_6,\lambda_7\neq 0$) can be generated radiatively.
The $\lambda_5$ generated
is typically very small, though, $\lambda_5\lsim 0.01$. 
In fact, the stability
of the 3d theory itself sets an upper bound: 
$\lambda_5 \gsim 0.03$ would 
make the theory unbounded from below.

For the mass parameters, 
there are essentially two conditions to be satisfied.
The first is rather easy to understand: denoting
$M_{12}^2 = m_{12}^2 + \fr12 \lambda_6 v_1^2 + \fr12 \lambda_7 v_2^2$,
we note that there is
a non-trivial minimum for $\phi$, if~\cite{lee} 
\be
\frac{|M_{12}^2|}{2\lambda_5 v_1v_2} < 1.
\la{CPmin}
\ee
Since $\lambda_5$, as well as $\lambda_6,\lambda_7$, 
are small, this essentially means that $m_{12}^2$ should 
be very small compared with $v^2=v_1^2+v_2^2$. 

The second condition is less obvious, but in fact 
a stronger one. Indeed, one should take into account
that $v_1,v_2$ in \eq\nr{CPmin}
are not free parameters but are determined 
by the equations of motion. As was argued in~\cite{cpown}, 
this leads at the tree-level to the constraint
\be
m_1^2+m_2^2 \lsim 2 \lambda_5 v^2.
\la{CPmin2}
\ee

The aim of this study is to see how these conditions get modified
by radiative corrections, first at 1-loop level, and then in the
full non-perturbative system. We can then compare, e.g.,  
with the values of $m_{12}^2, m_1^2+m_2^2$ allowed by the MSSM.

\section{1-LOOP RESULTS}

\begin{figure}[tb]

\vspace*{0.2cm}

\epsfxsize=6.3cm
\centerline{\epsffile{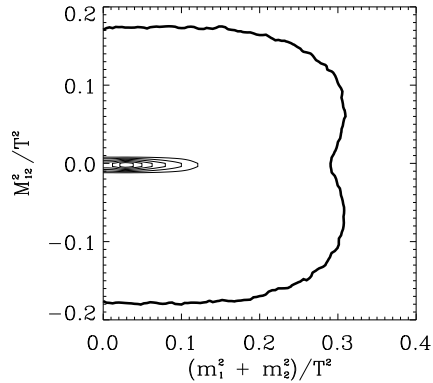}}
 
\vspace*{-1cm}

\caption[a]{A projection of the region of spontaneous
CP violation obtained at 1-loop level.} \la{scan}
\end{figure}

Adding to the tree-level potential the dominant 1-loop
corrections from the vector bosons, the parameter space 
leading to CP violation
can still be completely mapped out. We show a projection of
this region in \fig\ref{scan}. In this figure, 
we have imposed an upper bound $v/T \lsim 3$, since it is only
such values which are realized around the electroweak phase 
transition, and also since otherwise the finite temperature 
expansion used in the construction of the effective theory
breaks down. 

After the constraint on $v/T$, $\lambda_5$ is restricted
to small values, $\lambda_5\lsim 0.02$. We observe
that $M_{12}^2$ is small 
compared with $v^2$, as required by \eq\nr{CPmin}.  
The upper bound on $m_1^2+m_2^2$ given by \eq\nr{CPmin2}, 
on the other hand, gets 
modified by numerical factors, but not in order of magnitude. 

\section{COMPARISON WITH LATTICE}

\begin{figure}[tb]

\vspace*{0.2cm}

\epsfxsize=5.3cm
\centerline{\epsffile{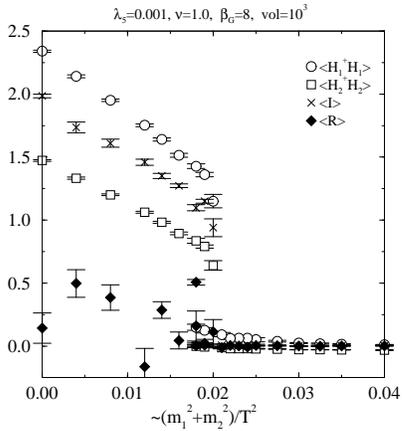}}
 
\vspace*{-1cm}

\caption[a]{Lattice
results for specific parameters.} \la{mflatt}
\end{figure}

Finally, we wish to compare the perturbative results 
in \fig\ref{scan} with non-perturbative lattice results. 
In order to do so, we choose a particular way of 
increasing $m_1^2+m_2^2$ in a way that we cross the 
boundary of the CP violating phase. We then wish to 
see whether larger values are allowed
in the full system than in \fig\ref{scan}. 

The operators we measure on the lattice are $H_a^\dagger H_a$ and 
$R = \re H_1^\dagger \tilde H_2, 
I = \im H_1^\dagger \tilde H_2$. 
Here $I$ is the order parameter for CP violation.

Typical results from lattice simulations
at small volumes are shown in \fig\ref{mflatt}.
We observe that the location of the transition 
does not change significantly from the
perturbative value $\sim 0.015$ for 
these parameters. We thus conclude 
that perturbation theory is relatively 
reliable, and \fig\ref{scan} is quite
a satisfactory approximation
for the part of the parameter space leading 
to CP violation. 

\section{IMPLICATIONS FOR THE MSSM}

We can now compare the region leading to
CP violation with that allowed by the finite $T$ MSSM. 
Consider, in particular, \eqs\nr{CPmin}, \nr{CPmin2}, 
or the corresponding projections in \fig\ref{scan}.

First of all, it turns out that
a small $m_{12}^2$ as required by \eq\nr{CPmin} can be 
obtained more easily at finite $T$ than at $T=0$. 
Thus, it seems that spontaneous CP violation could 
be possible. 

Unfortunately, the other constraint, \eq\nr{CPmin2}, 
is stronger and works in the opposite direction. 
Indeed, at finite $T$ in the MSSM
(see, e.g., \cite{cpown}), 
\be
m_1^2 + m_2^2 \approx m_A^2 +
0.53 T^2 +... \la{sum}
\ee
where the remainder is positive. 
Thus, at temperatures below $\sim 100$ GeV relevant
for the electroweak phase transition, $(m_1^2+m_2^2)/T^2\gsim 1.2$
for $m_A\gsim 80$ GeV. This does not agree with \eq\nr{CPmin2},
nor overlap with 
the region shown in \fig\ref{scan}.

\section{OUTLOOK}

We have addressed the first of the questions
of the Introduction, and found a negative
answer: one does not end up in a phase
where CP is spontaneously violated in the MSSM.
A light stop changes some aspects of 
the analysis, but the conclusion seems to remain 
the same. 

Let us then consider the second question.
Now, the analysis above indicated that
spontaneous CP violation is always more likely
at large vevs, cf.\ \eqs\nr{CPmin}, \nr{CPmin2}. 
Thus it seems quite unlikely that CP would be 
violated in the phase boundary, if it is not
violated in the broken phase. As we have 
found that the broken phase of the MSSM is 
always CP symmetric, the answer to the latter
question seems also to be negative. A solution 
of the perturbative equations of motion within 
the phase boundary points
in the same direction~\cite{pj}.

This issue, as well as other
related problems, can of course still be studied 
with lattice simulations. Work is in progress
(see also~\cite{zf}).



\end{document}